\documentclass{emulateapj}
\usepackage{amssymb,amsmath,epsfig,graphicx, natbib}

\def\sles{\lower2pt\hbox{$\buildrel {\scriptstyle <}
   \over {\scriptstyle\sim}$}} 
\def\sgreat{\lower2pt\hbox{$\buildrel {\scriptstyle >}
    \over {\scriptstyle\sim}$}} 

\def\kms{\mbox {~km~s$^{-1}$}}

\def\mpc3{\mbox {~Mpc$^{3}$}}

\def\asec{\mbox {\ifmmode {'' }\else $''~$\fi}}  
\def\amin{\mbox {\ifmmode {' }\else $'~$\fi}}    
\def\arcsper{\mbox {\ifmmode \rlap.{'' }\else $\rlap{.}'' $\fi}} 
\def\arcmper{\mbox {\ifmmode \rlap.{' }\else $\rlap{.}' $\fi}} 

\def\um{\mbox {${\mu \rm{m}}$}}
\def\deg{\mbox {$^{\circ}$}}
\def\msun{\mbox {${\rm ~M_\odot}$}}

\def\lsun{\mbox {${~\rm L_\odot}$}}
\def\msunyr{\mbox {$~{\rm M_\odot}$~yr$^{-1}$}}

\def\Pa{\mbox {Pa$\alpha$}}
\def\lya{\mbox {Ly$\alpha$}}
\def\Ha{\mbox {H$\alpha$}}

\def\h0{\mbox {~H$_0$}}
\def\q0{\mbox {~q$_0$}}
\def \i {\mbox{${\rm IRAS}$}}
\newcommand{\be}{\begin{equation}} \newcommand{\ba}{\begin{eqnarray}}
\newcommand{\ee}{\end{equation}} \newcommand{\ea}{\end{eqnarray}}

\def\-{{\em{---}}}

\def \h         {\hbox{$\, h$} }

\def\H7{\mbox {$h_{0.7}$}}
\def\naI{\mbox {\ion{Na}{1}}}

\def\IZw18{I~Zw~18}
\def\m82{M82}

\def\h{\mbox {\rm H}}

\def\deg{\mbox {$^{\circ}$}}
\def\msun{\mbox {${\rm ~M_\odot}$}}

\def\lsun{\mbox {${~\rm L_\odot}$}}
\def\msunyr{\mbox {$~{\rm M_\odot}$~yr$^{-1}$}}

\def\lya{\mbox {Ly$\alpha$}}

\def\Ha{\mbox {H$\alpha$}}

\def\Pa{\mbox {Pa$\alpha$}}

\def\h0{\mbox {~H$_0$}}
\def\q0{\mbox {~q$_0$}}
\def\o3hb{[OIII]$\lambda5007$~/~H$\beta$~}
\def\O1ha{[OI]$\lambda6300$~/~H$\alpha$~}

\def\s2ha{[SII]$\lambda\lambda6717,31$~/~H$\alpha$~}
\def\2z2{HeII~$\lambda4686$~}
\def\z7{[NII]~$\lambda6583$ }
\def\N2{[NII]~$\lambda6583$~/~H$\alpha$~}
\def\16z2{[SII]~$\lambda\lambda6717, 6731$ }

\def\asec{\ifmmode {'' }\else $''~$\fi}  
\def\amin{\ifmmode {' }\else $'~$\fi}    
\def\arcsper{\ifmmode \rlap.{'' }\else $\rlap{.}'' $\fi} 
\def\arcmper{\ifmmode \rlap.{' }\else $\rlap{.}' $\fi} 
\def\sles{\lower2pt\hbox{$\buildrel {\scriptstyle <}
   \over {\scriptstyle\sim}$}} 
\def\sgreat{\lower2pt\hbox{$\buildrel {\scriptstyle >}
    \over {\scriptstyle\sim}$}} 
\def\kms{\mbox {~km~s$^{-1}$}}

\def\cm3{~cm$^{-3}$}

\def\mpc3{~Mpc$^{3}$}
\def\mpc-3{~Mpc$^{-3}$}

\def\um{~${\mu}$m}

\def\Pa{\mbox {Pa$\alpha$}}
\def\clm{ }
\def\new{ }

\begin{document}

\title{Resolving Gas Flows in the Ultraluminous Starburst IRAS23365+3604 with Keck LGSAO/OSIRIS\footnotemark[$\dagger$]}
\footnotetext[$\dagger$]{The data presented herein were obtained at the W.M. Keck Observatory, which is operated 
as a scientific partnership among the California Institute of Technology, the University of California and the 
National Aeronautics and Space Administration. The Observatory was made possible by the generous 
financial support of the W.M. Keck Foundation. The data were obtained with the OH Supressing Infrared Spectrograph (OSIRIS)
behind the Laser Guide Star Adaptive Optics System}

\author{Crystal  L. Martin\altaffilmark{1} and Kurt T. Soto\altaffilmark{2}}
\altaffiltext{1}{Physics Department, University of California, Santa Barbara, CA 93106-9530}
\altaffiltext{2}{Institute for Astronomy, ETH Zurich, Zurich 8093, Switzerland}

\begin{abstract}
{\clm
Keck OSIRIS/LGSAO observations of the ultraluminous galaxy IRAS~23365+3604 resolve a circumnuclear bar 
(or irregular disk) of semimajor axis  0\farcs42 (520 pc) in \Pa\ emission.
The line-of-sight velocity of the ionized gas increases from the northeast towards the southwest;
this gradient is perpendicular to the photometric major axis of the infrared emission. 
Two pairs of bends in the zero-velocity line are detected. The inner bend provides evidence
for gas inflow onto the circumnuclear disk/bar structure. We interpret
the gas kinematics on kiloparsec scales in relation to the molecular gas disk and multiphase outflow 
discovered previously.}
 In particular, the fast component
of the ouflow (detected previously in line wings) is not detected, adding support to the conjecture that 
the fast wind originates well beyond the nucleus.  These data directly show the dynamics of gas inflow and outflow 
in the central kiloparsec of a late-stage, gas-rich merger and demonstrate the potential of integral field 
spectroscopy to improve our understanding of the role of gas flows during the growth phase of bulges 
and supermassive black holes.
\end{abstract}

\section{Introduction}
\label{sect:intro}

Gravitational torques generated during galaxy mergers provide a widely accepted mechanism to transport 
gas from the outer regions of galaxies inwards where it can fuel the growth of stellar bulges and
supermassive black holes \citep{Springel2005b,Hopkins2006a,Hopkins2010}. This growth is likely regulated by
feedback from the starburst and active galactic nucleus, or AGN \citep{Tremaine2002,
DiMatteo2005,DiMatteo2008}.
Accurately modeling these gas inflows and outflows remains central to understanding how gas
dynamical processes shape the structural properties of galaxies. Resolving these flows during 
the peak era of bulge formation at $z \sim 2$ is not practical, so we study local analogs 
in order to better understand the dominant physical processes.

Galaxies with exceptionally high central concentrations of star formation, gas, and dust
are rare in the local universe but have been identified by their high far-infrared 
luminosities, $L / \lsun\ \sgreat 10^{12}$ \citep{Soifer1986,Sanders1986,Solomon1997a, Sanders1988}. 
This population of ultraluminous infrared galaxies (ULIRGs) selects major mergers of gas
rich galaxies \citep{Borne2000}.  The ULIRGs include a range of evolutionary stages from 
well-separated double nuclei  to fully coalesced systems  
\citep{Murphy1996,Veilleux2002}. Towards later merger classes, 
the average contribution of nuclear activity to the bolometric luminosity increases 
\citep{Veilleux2009}, supporting the long-standing conjecture that an
active nucleus emerges fairly late in the evolutionary progression \citep{Sanders1988b}.

Spatially resolved spectroscopy of ULIRGs provides strong evidence for inflowing
gas on scales from roughly 1~kpc to several tens of kpc.
For example, the increasing strength of Balmer absorption lines with galactocentric radius \citep{Soto2010}
reveals stellar age gradients which are inverted relative to the typical inside-out growth of disks \citep{Larson1976,
Naab2006,Matteucci1989,MacArthur2004,Boissier2008}. The relatively young ages in the central few kpc's
of ULIRGs and the paucity of young stars at large radii implies that star formation is truncated in
the outer disk several hundred Myr before the gas supply is depleted in the central kpc.
Also, in contrast to the normal decline of metallicity with radius in galaxies, ULIRGs have
shallower or even 
{\clm inverted metallicity gradients \citep{Rupke2008,Rich2012}.}
 Between first and second pericenter
passage, inflow of lower metallicity gas from large radii dilutes the central metallicity.
Inflows therefore clearly transport gas from the outer disk into the central kpc of the merger remnant.

{\clm
In numerical simulations of major mergers, the high central gas densities trigger 
gravitational instabilities forming features like bars, rings, and spirals on sub-kiloparsec
scales. An accurate description of these gas inflows on circumnuclear scales requires 
proper treatment of the physics on considerably larger spatial scales because the gas fueling is 
typically driven by  kiloparsec-scale features, and the formation and evolution of these
kiloparsec-scale features is governed by the dynamics of the galactic disks on scales of order 10 kpc
\citep{Chapon2013,Emsellem2015}.  The hydrodynamic interaction with gas outflows driven
by supernovae and/or stellar radiation compounds this computational challenge but may
significantly delay the decay of the two supermassive black holes (SMBHs)
from separations of a few hundred parsecs to a few parsecs \citep{Roskar2015}. Fueling
by gravitational instability naturally delays the peak activity from the SMBH until 
star formation and feedback substantially reduce the gas fraction, a timescale of order
$10^8$~yr on kiloparsec scales \citep{Hopkins2012}. Testable predictions of these theoretical models
include (1) the identification of shocks on sub-kpc scales (via morphological features like bars, rings, and spirals 
for example), (2) the  properties of gas outflows from these regions,  and  (3) their temporal evolution during the merger
progression.
}

Directly observing this inflow on sub-kpc scales has only recently become possible.
Using the Keck I and II adaptive optics (AO) systems to obtain near diffraction-limited data cubes, \cite{Medling2014} 
found that nuclear disks were common in those (U)LIRGS classified as late stage mergers. Stellar nuclear disks are young
and indicate recent inflow. The gaseous component of nuclear disks plays an important role in angular momentum transfer
and is critical for the coalesence of binary black holes \citep{Chapon2013,Cole2014,Roskar2015}.
{\new
Circumuclear disks have effective radii of a few hundred parsecs and masses between $10^8$ and 
$10^{10}$\msun\ \citep{Medling2014}.
}
The formation of this dynamically cold component of gas and stars in the nucleus appears to be required for 
triggering Seyfert activity \citep{Hicks2013}. 
{\clm
Nuclear spirals, which are shocks in a circumnuclear disk, appear to
be a common mechanism for feeding in the gas \citep{Martini1999,Maciejewski2004a}.} 


Resolving scales of roughly 100~pc in 
both active galaxies \citep{Muller2011,Davies2014}
and ULIRGs \citep{Cazzoli2014,Medling2015,GarciaBurillo2015} 
has  provided new insight about gas transport out of the central region.  \cite{Muller2011} 
resolved emission from the narrow line region (NLR) and coronal line region of nearby AGN; the velocity fields
of the NLR showed both disk rotation and bipolar outflows.
In a study of a nearby buried quasar, F08572+3915:NW, \cite{Rupke2013b} 
found a close correspondence between the velocities of the blueshifted $H_2(1-0)$ emission component mapped with OSIRIS 
and the Herschel OH line profile of \cite{Veilleux2013}, whereas optical emission and absorption lines traced
a different part of the wind.

Galactic winds are ubiquitous among ULIRGS \citep{Martin2005,Rupke2005b,Soto2012b,Bellochhi2013,Veilleux2013,Rich2014,Cicone2014}.
The physical relationship between the outflowing gas mapped over scales of many kpc \citep{Martin2006,Rupke2015} and
the activity in the circumnuclear region of ULIRGs, however, remains relatively unexplored. 
Connecting high-resolution data cubes with the
global gas kinematics may prove particularly interesting for those ULIRGs with very broad, blueshifted emission-line
wings. Detections of similar features in spectra of high-redshift galaxies have recently been attributed to
AGN \citep{Genzel2014}, whereas the strength and profile of the wings in \lya\ spectra of cool ULIRGs is more simply
understood in the physical context of gas condensing out of the hot phase of winds \citep{Martin2015,Thompson2015}.
In the latter scenario, the highly blueshifted ($|v| > 500$\kms) gas cools at radii of a few kpc, whereas
AGN outflows are accelerated on much smaller spatial scales. The velocities of molecular outflows have been
shown to increase with the luminosity of the active nucleus \citep{Veilleux2013}, but understanding whether the
molecular gas is accelerated near the nucleus or forms in situ in the cooling wind at much larger radii requires
spatial mapping. 

In this paper, we seek to combine the resolution offered by adaptive optics with information from the wider
fields obtained with seeing-limited observations and thereby identify signs of gas flows over several
decades in radius. We present the gas kinematics and continuum morphology in the central region of
IRAS 23365+3604, a ULIRG with a low central metallicity \citep{Rupke2008} and inverted stellar age gradient \citep{Soto2010},
at roughly $100$ pc resolution. 
{\clm
Archival images from the Hubble Space Telescope (HST, PROPOSID 10592 and 6346)  facilitate identification of
a suitable tip-tilt star for the AO system, enable accurate positioning of the OSIRIS (OH-Suppressing
Infra-Red Imaging Spectrograph, \cite{Larkin2006a}) lenslet array, and allow construction of a high-resolution
color map.}
The redshift of the system shifts the
Paschen-$\alpha$ 1875.13~nm emission into a relatively clean region of the airglow spectrum in the $K$ band
where we observed.

In \S 2 we describe the new Keck/OSIRIS observation. Measurements of the continuum morphology
and the \Pa\ morphology and kinematics  are described in \S 3. 
We argue that the overall velocity field
reveals an outflow launched from a circumnuclear disk, and that the disk shows signs of gas inflow.
In \S~\ref{sec:discussion}, we discuss the physical relationship of the
associated gas flows to previous observations of this system on larger spatial scales.
In \S~\ref{sec:summary}, we summarize our main results.
Throughout this paper we adopt a $\Lambda$CDM cosmology with $H_0 = 70$\kms Mpc$^{-1}$, $\Omega_m = 0.30$,
and $\Omega_{\Lambda} = 0.70$. We adopt the CO redshift, $z = 0.064480$ \citep{Solomon1997a}, which gives an angular
scale of 1.239~$h_{70}^{-1}$~kpc per arcsecond. Throughout the manuscript we define the position angle (PA)
on the sky as the angle measured eastward from north.

\section{Data}
\label{sect:obs}

\begin{figure}[h] 
\centering
\includegraphics[width=0.85\linewidth, keepaspectratio=true, angle=0, clip=true, trim = 200 0 0 0]{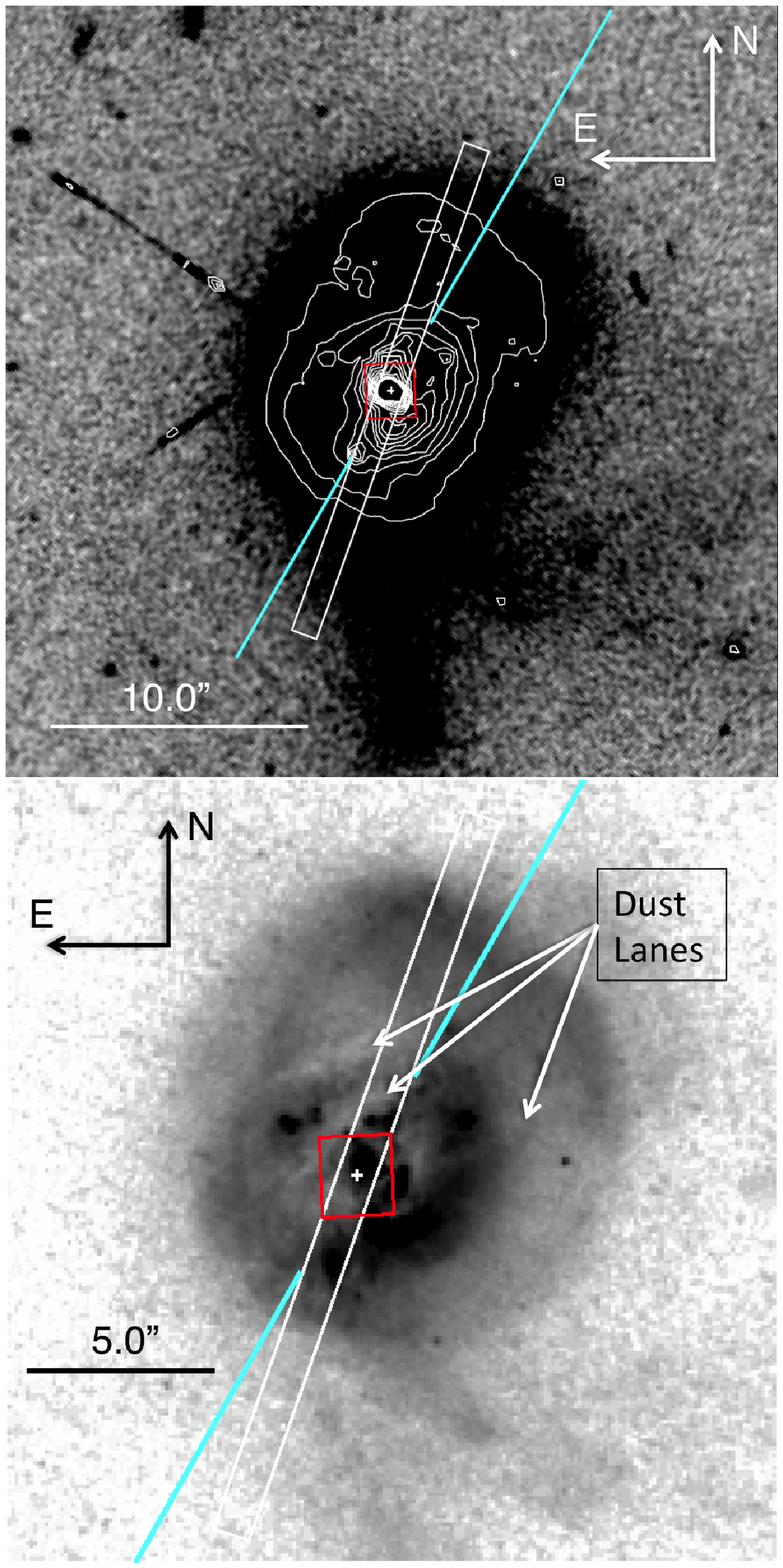}
\caption{{\it Top:}
HST F814W image of IRAS 23365+3604 with contours showing isophotes.
The cyan line marks the major axis of the galaxy at $PA = -30\deg$.
The red box indicates the $2\farcs10 \times 2\farcs24$ region mapped with OSIRIS.
The white rectangle indicates the position of the 
 slit used to obtain the 
echellete spectrum previously discussed by \cite{Martin2005}, \cite{Martin2006}, \cite{Soto2010}, 
\cite{Soto2012a}, and \cite{Soto2012b}.
{\it Bottom:}
{\clm
Map of $B-I$ color constructed from archival ACS/F435W and WFPC2/F814W 
HST observations. Two prominent dust lanes (white) are visible  1\farcs5 and 3\farcs0 northeast of the nucleus; 
a fainter dust lane approaches the southwest corner of the OSIRIS map (red box). 
}
}
\label{fig:zoom} \end{figure}

The HST image and color map in Fig.~\ref{fig:zoom} illustrates the environment of the region mapped in \Pa. 
Tidal features are plainly visible to the south and southeast; a deep
stretch of the contrast shows additional tidal features to the northeast.
The stellar population has already aged several hundred
Myr in much of the outer galaxy \citep{Soto2010}. Along the ESI slit shown, \Ha\ emission is
detected from the tidal feature located roughly 6\arcsec\ north of the nucleus and from
the southern tidal arm. This tidal, ionized gas has  a velocity gradient opposite in sign to the shear along the 
inner contours, and this counter rotation is shared by the molecular gas on comparably large
spatial scales (see position-velocity diagram in Fig. 3 of \cite{Cicone2014}). 

Table~\ref{tab:galprop} summarizes the properties of IRAS 23365+2604.
The high $L_{FIR}$ indicates a star formation rate (SFR) over a hundred \msunyr. At this rate, the
molecular gas supply will be exhausted on a timescale of 100~Myr, so we are clearly observing an object
undergoing a rapid transition towards a more quiescent state.

\begin{deluxetable}{lc} 
\tablewidth{0pt}
\tablecaption{IRAS~23365+3604 \label{tab:galprop}} 
\tabletypesize{\small}
\tablehead{
\colhead{Property}	&
\colhead{Value}		
}
\startdata
Redshift\tablenotemark{a}                    & 0.064480 \\
$\log ( \rm L_{IR}/L_{\odot}$)\tablenotemark{b} & 12.20 \\
$\log (\rm L_{FIR}/L_{\odot}$)\tablenotemark{c}& 11.96 \\
SFR (M$_{\odot}$  yr$^{-1}$)\tablenotemark{d}& 280 \\
F$_\nu ($12$\mu$m) (Jy)                      & $ \le\ 0.25$ \\
F$_\nu ($25$\mu$m) (Jy)                      & 0.81 \\
F$_\nu ($60$\mu$m) (Jy)                      & 7.69 \\
F$_\nu ($100$\mu$m) (Jy)                     & 8.19 \\
W(\naI) (\AA)\tablenotemark{e}               & $4.5 \pm 0.4$ \\
$V_{max}$ (\naI) (\kms)\tablenotemark{f}    & -651 \\
$V_{B}$ (\naI) (\kms)\tablenotemark{g}       & $-389 \pm 72$ \\
$\Delta V_{B}$ (\naI) (\kms)\tablenotemark{h} & $-308 \pm 85$ 
\enddata
\tablenotetext{a}{The redshift derived from a CO (1-0) observations \citep{Solomon1997a}}
\tablenotetext{b}{Estimated $8 – 1000$ $\mu$m  luminosity.
The IRAS 12 $\mu$m flux is an upper limits, so the median flux 
ratios of bright ULIRGs with IRAS detections in all four 
bands were used to estimate $L_{IR}$ as described in \cite{Murphy1996}.}
\tablenotetext{c}{Far-infrared luminosity computed from $L_{FIR}
= 3.86 \times 10^5 d_L^2 [2.58 F_{\nu}(50 \mu m) + F_{\nu}(100 \mu m)] L_{\odot}$,
where the flux density is in Janskys and the luminosity distance is in Mpc.}
\tablenotetext{d}{Star formation rate estimated from $SFR (\msunyr) =  L_{IR} / 5.8 \times
 10^9$ L$_{\odot}$ \cite{Kennicutt1989}. This relation assumes continuous star formation and a 
Salpeter initial mass function from 0.1 to 100 M$_{\odot}$.}
\tablenotetext{e}{Equivalent width of interstellar resonance absorption in \naI\ $\lambda
\lambda 5890, 96 $ \citep{Martin2005}.}
\tablenotetext{f}{The maximum blueshift of the \naI\ absorption trough \citep{Martin2005}.}
\tablenotetext{g}{The blueshift of the fit to the interstellar, \naI\  absorption trough \citep{Martin2005}.}
\tablenotetext{h}{The width of the interstellar \naI\  absorption line \citep{Martin2005}.}
\end{deluxetable}

\subsection{Observations}

We observed IRAS 23365+3604 at Keck II on September 3, 2010 using the OSIRIS grating in third order with the Kn1 filter 
and the laser guide star adaptive optics (LGS-AO) system \citep{Wizinowich2006} as described by Table~\ref{tab:observations}.
Conditions were clear; and the atmospheric seeing improved slowly over the course of the night from an initial value of
approximately 0\farcs80 FWHM. We configured OSIRIS with the 0\farcs035 lenslet array which slightly undersamples
the diffraction limited core of the point spread function (PSF)  but enlarges the field of view of the $36 \times 64$ lenslet 
array to 1\farcs26 $\times$ 2\farcs24 per pointing. We obtained 2292 spectra at each pointing covering the complete spectral 
range from 1955 - 2055~nm at 0.25 nm per pixel. 

\begin{deluxetable*}{lllllllll}[b] 
\tablewidth{0pt}
\tablecaption{ Observations \label{tab:observations}}
\tabletypesize{\small}
\tablehead{
  Name			&  UT Date	& $t_{\rm exp}$	&$n_{\rm exp}$	& $t_{\rm tot}$	& offset  	& PA		& Filter	& Scale	\\
  (1)				&  (2)		& (3)      		& (4)			& (5)			& (6)	        	& (7)		& (8)		& (9)		
}                                                                                                           
\startdata                                                                                                  
23365$+$3604:West	&  09-03-2010	& 900		& 8			& 7200         	& 0.5	  	& 272	& Kn1	& 0.035	\\ 
23365$+$3604:East	&  09-03-2010	& 750		& 7			& 5250         	& -0.5	& 272	& Kn1	& 0.035	
\enddata
\tablecomments{ \scriptsize
Col.(1): IRAS name:pointing.
Col.(2): Universal time observation date.
Col.(3): Exposure time per dither/sky location in seconds.
Col.(4): Number of target exposures.
Col.(5): Total target exposure time in seconds
Col.(6): Pointing offset from the nucleus of the galaxy. Each exposure was
			dithered in a box pattern by 0\farcs035 in the lenslet array x and y coordinate system
			around this central location.
Col.(7): Position angle of the lenslet array in degrees.
Col.(8): Spectrograph filter.
Col.(9): Spatial scale of the lenslet array in units of arcseconds
}
\end{deluxetable*}

We established the offset between the field of view of this Kn1 configuration and the center of the instrument 
field of view by observing a bright star with the Kn3 filter (0\farcs020 scale) and comparing this to the same object 
acquired in the Kn1 configuration (and matched 0\farcs020 scale). Using this offset we acquired the tip tilt star in 
spectroscopic mode using the Kn1 filter (0\farcs035 scale), offset to the galaxy position defined by the nucleus in the 
F814W image, and tweaked the position to center the object in the Kn1 filter.

Using two pointings offset along the shorter dimension of the lenslet array, we mapped a 2\farcs10 $\times$ 2\farcs28 
region ($2.60 {\rm ~kpc~} \times 2.83 {\rm ~kpc~}$) centered on the brightest continuum knot. At each position, we 
dithered exposures in a box pattern with step size 0\farcs070 along the axes of the lenslet array.
The individual exposure times were 900 and 750 s, respectively, for the eight western and seven eastern positions.
Since the entire field of view is covered by the object, leaving no empty sky spaxels, we employed an object/sky/object 
nodding pattern (of a few arc minutes). 

The LGS-AO system produced an artificial guide star directly on the OSIRIS optical axis. 
We locked the tip-tilt sensor onto a natural guide star of magnitude $r = 16.7$ at an angular separation of
55\farcs2 from the nucleus and observed with a closed AO feedback loop. 
Between the last exposure at the western pointing and the first exposure at the eastern pointing,
we observed a pair of stars with similar angular separation and PA as  IRAS 23365+3604 
and 
{\bf its}
 tip-tilt star in order to characterize the PSF.
We also performed repeated observations of two A0 stars at similar airmass as the target.



\subsection{Data Reduction}


{\clm 
Individual exposures were first reduced using Version 2.3 of
the OSIRIS data reduction pipeline \citep{Larkin2006a}.
}
The presence of the bright galaxy, which covered a significant fraction of the detector, skewed
the pipeline calculation of the offset level in each channel.\footnote{The detector is read out in 32 channels. 
         Each region has a slightly different zero-level intensity.}
To circumvent this problem, we identified the subregions of each channel least affected by the galaxy and calculated 
the offset level using iterative sigma  clipping. We treated these offsets as bias levels and subtracted them from
each channel prior to re-processing with the pipeline. The pipeline processing included steps for sky subtraction, 
cosmic ray removal, spectral extraction, dispersion correction, telluric feature removal, and datacube construction.
Observations of HIP114714 and HIP1603 were used to remove telluric features from the data cubes obtained
at the western and eastern pointings, respectively.

We registered the individual datacubes produced by the pipeline using the telescope offsets commanded between the pointings and 
dither positions; these shifts were, by design, an integer number of spaxels for the PA of the lenslet array.
To flag voxels (volume elements) with non-physical values in each cube, we calculated the median intensity and standard deviation 
among the individual cubes at each voxel. We constructed the final datacube from the aligned cubes using a single
iteration of a $100\sigma$ mean clip to reject unphysical voxels (typically bad lenslets).  We adopted the standard deviation 
among the individual datacubes, which was slightly larger than the error estimates produced by the pipeline, 
as the best estimate of the uncertainty in the intensity in each voxel.

\subsubsection{Dispersion Solution and Spectral Resolution}

To check the dispersion solution, we measured the centroids of five OH lines in the final datacube.
The measured lines include the OH transitions  near the observed $Pa\alpha$ emission
at vacuum wavelengths of 2000.8163~nm and 2003.3211~nm \citep{Rousselot2000}. 
{\clm
We found that the pipeline solution was too red and applied an offset of
-0.527 nm to the dispersion solution.\footnote{This bug was fixed in a later version of the pipeline.}
}
We computed and applied a heliocentric correction of 13\kms.

The spectral resolution has significant variation between lenslets and at different
wavelengths. We measure a median linewidth of 100\kms\ near the observed wavelength of \Pa,
consistent with the expected value of $R = 3100$ (cf., OSIRIS User Manual Figure 2-7).

\begin{deluxetable*}{lcccccccc}[b]
\tablewidth{0pt}
\tablecaption{ \label{tab:psfstars}PSF Stars}
\tabletypesize{\small}
\tablehead{
  type	& name & RA & Dec & b-r & b-v & r & $\Delta\theta$ & PA	\\
  (1)				&  (2)		& (3)      		& (4)			& (5)			& (6)	        	& (7)		& (8)		& (9)		
}                                                                                                           
\startdata                                                                                                  
tip tilt & 23385679+3621232 & 23 38 56.80 & +36 21 23.26 & 0.5 & 0.28 &16.7 & 55.9 & 210 \\
{\clm 
PSF tip tilt} & 23301627+3648251& 23 30 16.27 & +36 48 25.11 & 0.4 & 0.22 & 16.4 & 56.45 & 210 \\
PSF & 23302096+3648229 & 23 30 20.97 & +36 48 22.90 & 0.6 & 0.33 & 16.7
\enddata
\tablecomments{ \scriptsize
Col.(1): The type of calibration that the star will be used for.
Col.(2): 2mass PSC name
Col.(3): Right Ascension
Col.(4): Declination
Col.(5): b-r color
Col.(6): b-v color
Col.(7): r band magnitude
Col.(8): Angular separation between the target and the tip tilt star
Col.(9): Position Angle of the tip tilt star relative to the target.
}
\end{deluxetable*}

\subsubsection{PSF Estimation}

Knowledge of the PSF is required to model how observed spatial structure relates to the intrinsic morphology.
Since the wavefront reference for our observations is not the science target, constructing the PSF
from the wavefront reference would be misleading due to anisoplanaticism. Instead, we observed a pair 
of stars, the "PSF pair", near our field (within $15 \deg$) from  the 2MASS Point Source Catalog; 
{\clm this pair has a}
 separation (within 5\farcs0) and PA (within $15\deg$) very similar to that of IRAS 23365+3604 and its tip-tilt star 
(see Table~\ref{tab:psfstars}). The ``PSF tip-tilt star'' plays the same role as the tip-tilt star used 
in the galaxy observation. It provides low-order corrections for the LGS system and must have a similar 
color and magnitude to that of the science tip-tilt star to ensure a similar distribution of flux on the
wavefront sensor during the PSF-pair and science observations.  Table \ref{tab:observations} illustrates
the close match in color and magnitude of the two tip-tilt stars.

\begin{figure}[h]
\centering
\epsfig{file=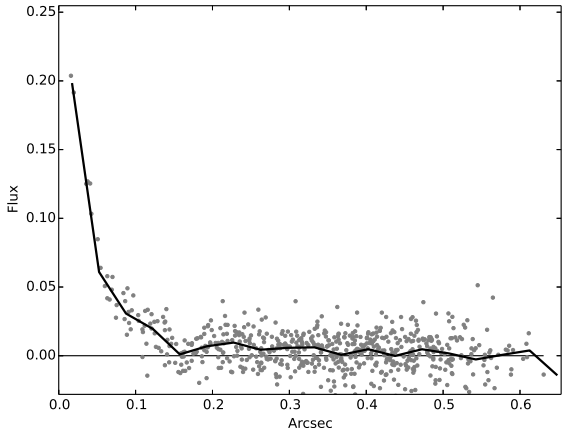, width=\linewidth, keepaspectratio=true}
\caption{
 The radial profile of the PSF star.  The gray points are the individual pixels of the PSF image. The black 
line is a median of azimuthal bins. The bin width,  0\farcs035, is matched to the lenslet scale. The core of the profile has a HWHM 
of 0\farcs045.}
\label{fig:psf}  \end{figure}

We expect the AO system to deliver a  PSF with a narrow core and broad wings \citep{Davies2007}. The diffraction
limit of the telescope and detector pixel sampling shape the core. Natural seeing determines width of the halo.
Figure~\ref{fig:psf} shows the PSF of the PSF star. 
A  Moffat function fitted to the profile yields a core of width 0\farcs090 FWHM wide.
The median profile plausibly shows  wings over the scale of the natural seeing. 
We avoid any direct estimate of the Strehl ratio because it is sensitive to errors in 
background level; and variations in the sky intensity between our object and sky frames 
produce small errors in the background that are difficult to quantify. 


In our analysis of IRAS 23365+3604, we model the AO PSF with a Gaussian function of 
FWHM matched to the spatial profile of the PSF star. This PSF is convolved with 
each intrinsic model before comparing the model to the observed datacube. This approximation
for the width of the core of the PSF suffices because it is inevitably the simplicity of our disk model 
that limits how well the convolved model matches the observed data cube.

\section{Results}
\label{sec:results}

\begin{figure*}
\centering
\includegraphics[width=0.65\linewidth, keepaspectratio=true, angle=90,clip=true, trim=100 0 0 0]{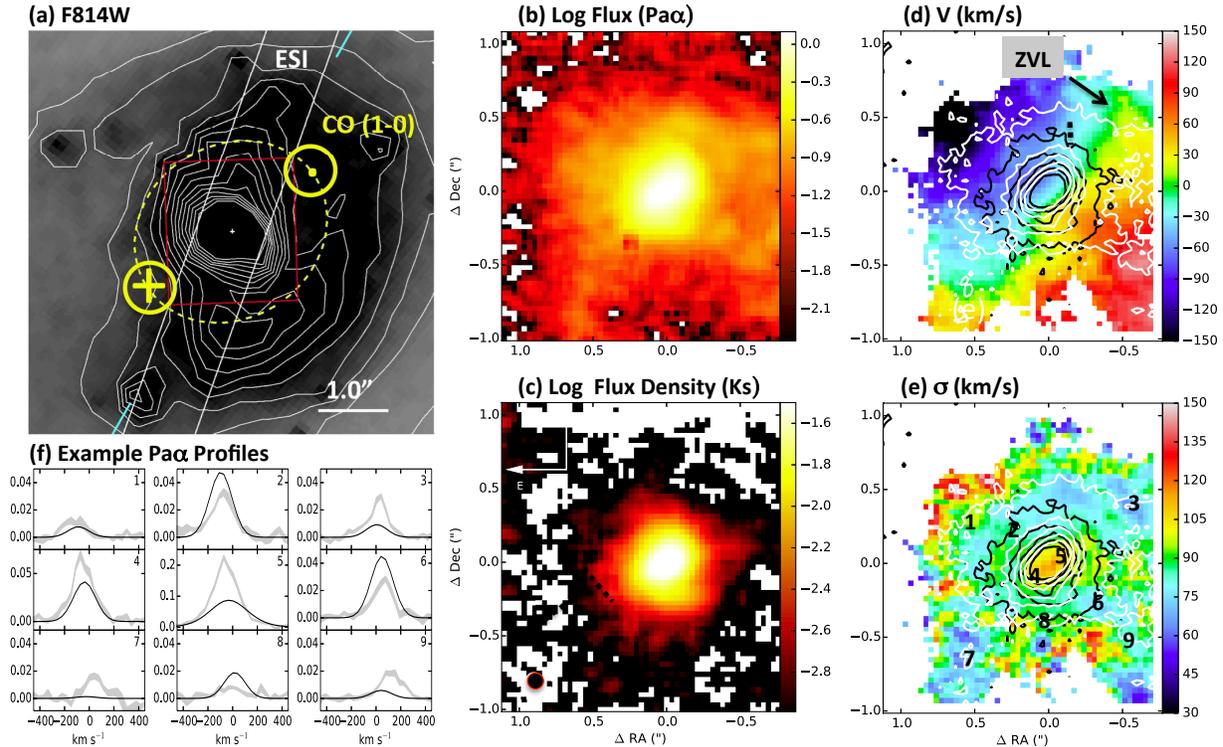}
\caption{\label{fig:fullmaps} 
{\clm
{\it (a)} 
HST F814W image from Fig.~\ref{fig:zoom} zoomed in on the central 6\farcs0 by 6\farcs0 
(7.4 kpc $ \times$ 7.4 kpc) region of IRAS 23365+3604. 
Cyan ticks define the galactic major axis as  in Fig.~\ref{fig:zoom}.
The yellow ellipse indicates the beam size of the CO (1-0) observation of \cite{Cicone2014} 
who identify a disk of molecular gas with line of nodes at $PA = -55\deg$;
the northwest side is approaching, and the southwest side receding. Red box indicates
the size of the 2\farcs10 $\times$ 1\farcs90 field of the OSIRIS map.
{\it (b,c)}
The 
{\new
smoothed 
}
\Pa\ line image and  the line-free, K-band continuum, respectively 
The origin is the infrared nucleus; the circle in the lower left corner of Panel~c 
indicates the resolution.
The flux scales are logarithmic with arbitrary zero points, and the regions
with fluxes smaller than the $1 \sigma$ flux error are masked.
{\it (d)} Doppler shift of \Pa\ emission.
The gradient in the line-of-sight velocity is along $PA = -34\deg$ and perpendicular to the contours of 
infrared  continuum (black) and line (white) emission. The surface brightness contours have been drawn at
90\%, 70\%, 50\%, 30\%, and 10\% of their peak values.   The colorbar shows the 
zero velocity line in green,  and we have masked spaxels with uncertainties greater than 10\kms.
See text for further discussion of the offset of the kinematic center from the position of the
infrared nucleus.
{\it (e)}
Velocity dispersion, $\sigma (\kms)$, of the \Pa\ emission. For comparison, 
the instrumental linewidth of 100\kms\ FWHM corresponds to  $\sigma = 42\kms$ and is 
indicated by the dark blue shading. Numbers show the locations of the example line profiles.
{\it (f)} Nine examples of  \Pa\ line profiles extracted from the data cube (gray) overlaid 
with the model (black) from Sec.~\ref{sec:bar}.
Comparison of the profiles at positions 
4 and 5 across the central, non-axisymmetric structure show no velocity offset, consistent
with their placement along the  direction of the ZVL. Line profiles extracted near the prominent filaments, 
labeled 1, 3, 7, and 9, confirm both the overall sense of the velocity gradient across the OSIRIS map and
the broader width of the observed lines (relative to the disk model).
}
}
\end{figure*}

The aim of this work is to relate the gas kinematics in the nuclear region
of a ULIRG to the galaxy as a whole. To provide context for the OSIRIS data cube,
we show an i-band image of the region surrounding the OSIRIS map in Figure~\ref{fig:fullmaps}.
The i-band morphology is complicated due to both the recent merger and spatial
variations in extinction. Across the region mapped with OSIRIS, the i-band
surface brightness contours are elongated northeast - southwest at $ PA = 50\deg$.
This elongation of the distribution of stellar light
contrasts sharply with the distribution of molecular gas.
\cite{Cicone2014} and \cite{Solomon1997a} derived position angles
of -55\deg and -45\deg, respectively, for the major axis of the molecular gas emission.
The \cite{Cicone2014} CO(1-0) map obtained with the IRAM PdB interferometer covers a  
field ten times larger than our OSIRIS map.
We have sketched the beam size of \cite{Cicone2014}'s CO (1-0) observation in Panel 
(a) of Figure~\ref{fig:fullmaps} to emphasize that the molecular gas measurements
do not resolve the gas kinematics within the OSIRIS map.

\subsection{Description of the Morphology}  \label{sec:data}

{\clm
Panels (b) and (c) of Fig.~\ref{fig:fullmaps} show the smoothed  OSIRIS data cube.
We experimented with several adaptive smoothing algorithms but show the result
of smoothing over each $3 \times 3$ block of spaxels at each wavelength slice.
The smoothing scale of 0\farcs105 maintains useful resolution in the circumnuclear
region while improving the signal-to-noise ratio of the \Pa\ nebula in the
central kiloparsec.}
Using this smoothed data cube, we fit a first-order polynomial to the 
continuum level on either side of the line and a single Gaussian line profile. 
The fitted continuum level defines the line-free, K-band, continuum intensity,
and the fitted flux defines the intensity of the $Pa\alpha$ image.

\subsubsection{Continuum Emission} 

The continuum image shows a single nucleus.
The FWHM of the point spread function in the OSIRIS image is 0\farcs091, so we estimate a maximum 
nuclear separation of 1.22 $\times\ FWHM$, a physical length of 138~pc. As suggested by 
the single nucleus morphology in seeing-limited K-band and R-band images, this galaxy is at a 
late stage of merging, after the nuclei have coalesced \citep{Murphy1996, Soto2010, Zamojski2011}. 
It is a Class~IV Merger in the system described by \cite{Veilleux2002}.\footnote{We note that \cite{Soto2012a}
   list this object as Class~IIIb in their Table~1 based on spatial information in an optical spectrum, but
   the absence of two nuclei in the higher resolution imaging presented here is inconsistent with
   a Class~III Pre-Merger stage.}

The K-band continuum isophotes are shown by black contours in Panels (d) and (e) of Fig.~\ref{fig:fullmaps}.
The major axis runs northwest -- southeast at $PA = -46\deg$, which is
remarkably well aligned with the galaxy major axis on large scales (indicated in Fig.~\ref{fig:zoom}).
The K-band isophotes are clearly not showing the same features as the F814W contours which run nearly
perpendicular to them. We attribute the very different morphologies of the optical and infrared images
to the extremely high extinction in the ULIRG.

We fit the K-band image with Sersic profiles of fixed index $n$. Each parametric model was convolved 
with a Moffat profile of 0\farcs09 FWHM (to approximate the OSIRIS psf) before comparing to the data.
The continuum surface brightness at radii 
$R \le\ 0\farcs5$ is  well fitted by the exponential model ($n=1$) with radial scalelength 
$R_e = 0\farcs130$ (161~pc). The surface brightness profile flattens at larger radii. 
The break in slope indicates a distinct morphological component that defines the circumnuclear region.

The size of this circumnuclear component is similar to that of nuclear disks. \cite{Medling2014},
for example, identified nuclear disks in nearby (U)LIRGs and measured effective radii of a few 
hundred parsecs. If the circumnuclear structure in IRAS~23365+3604 is intrinsically round, then the ratio
of the semi-minor to semi-major axis implies a disk inclination of approximately 38\deg. 
\cite{Medling2014} measured rotation in both the stellar and gaseous components of the nuclear
disks. We will examine the kinematics of the ionized gas in the circumnuclear structure 
in Sec.~\ref{sec:data_kinematics} below.

\subsubsection{Morphology of Ionized Gas Emission}  \label{sec:morph_gas}

Like the K-band continuum isophotes, the highest surface brightness \Pa\ isophotes are elongated in the 
southeast -- northwest direction. The contours, shown by white isophotes in Fig.~\ref{fig:fullmaps},
have nearly the same position angle and centroid as the continuum emission.  The \Pa\ emission is
detected at high S/N ratio to larger radii than the continuum emission. 
The \Pa\ isophotes become nearly circular  as the radius increases from 
0\farcs224 ($R = 278$~pc) to roughly 0\farcs350 ($R \approx 434$~pc).  
This change in the ellipticity indicates the image of the ionized gas may be a 
superposition of multiple structures.

To characterize the size scale of these structures, we fit concentric ellipses to the image and produced
a radial \Pa\ surface brightness profile. We obtained a good fit with an exponential
function on scales  $R < 0\farcs42$ (520~pc); the scale length ($R_e = 0\farcs200$ or 248 pc) was
slightly 
larger than measured for 
{\new
the stellar circumnuclear disk.}
 The surface brightness profile flattens 
from $R = 0\farcs42$ (520~pc) to $R = 1\farcs00$ (1240~pc),
and we fitted a scalelength $R_e = 0\farcs34$ (421 pc).
The source of this flattening is the lower surface brightness \Pa\ emission, clearly visible 
on scales $R > 0\farcs35$ in Fig.~\ref{fig:fullmaps}. 

We will refer to this lower surface brightness emission as the {\it extended emission} in
contrast to the higher surface brightness {\it circumnuclear emission} at $R < 0\farcs42$.
The extended \Pa\ emission at $PA \approx -90\deg$ is clearly visible in Fig.~\ref{fig:fullmaps}b
over a kiloparsec to the western edge of the map. We will produce a sharper view of the
filamentary structures when we subtract a smooth \Pa\ emission component in Sec.~\ref{sec:bar}.

\begin{figure} 
\centering
\includegraphics[width=\linewidth, keepaspectratio=true, angle=90,clip=true, trim=0 0 0 0]{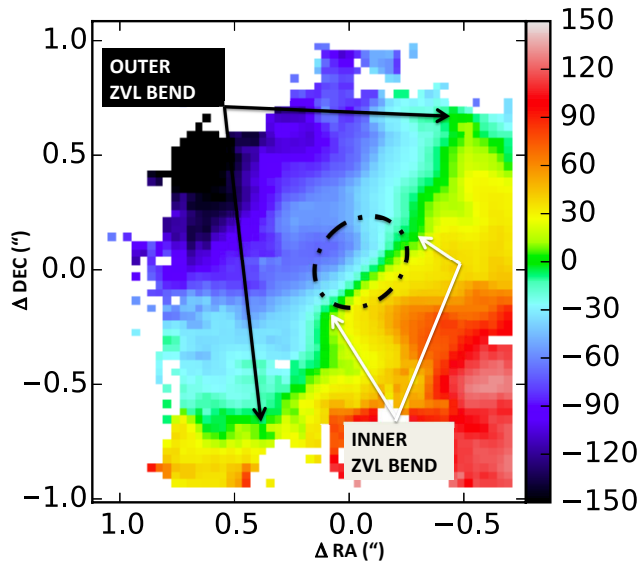}
\caption{
{\clm
Two pairs of bends in the zero velocity line are highlighted by the green contour in the \Pa\ velocity map. 
Gas on circular orbits in a disk  would produce a straight ZVL. When the velocity field of the gas includes
a radial inflow component, as in barred galaxies, kinks in the ZVL as seen here are often detected.  
In \i23365+3604, the inner bend is located at the break in the infrared surface brightness which defines the 
circumnuclear region (dot-dashed ellipse with semi-major axis of 520 pc). 
The outer pair of kinks in the ZVL are located 0\farcs79 (0.97 kpc) from the nucleus
where the ZVL makes a right angle turn clockwise,  takes on a $PA \approx 44\deg$, and then reaches the 
outer edge of the OSIRIS map.}
}
\label{fig:model_bar} \end{figure}

\subsection{Description of the \Pa\ Gas Kinematics}  \label{sec:data_kinematics}

{\clm
We adopt the CO (1-0) redshift of $ z = 0.064480$ from \cite{Solomon1997a}
to define the systemic velocity. Since this observation is integrated over
the entire galaxy, we expect it to indicate the true barycenter of the ULIRG. 
We assign a systematic uncertainty of 17\kms\  to this redshift, however,
based on an unpublished redshift of $z= 0.064419$  kindly derived for us 
from another CO data cube (C. Cicone, pvt. comm.)}

{\clm
The central wavelength of our integrated \Pa\ profile agrees with the near-infrared 
spectrum presented by \cite{Murphy2001}. With the above  definition of the systemic
velocity, the integrated \Pa\ line profile is blueshifted 30\kms.  The net blueshift
would be reduced to 13\kms\ had we adopted the $z= 0.064419$ redshift,
so we claim a net blueshift of  $13 - 30$\kms\ for the ionized gas in the central
kiloparsec  relative to the molecular gas on galactic scales.
The nucleus is not optically thin in \Pa, and we interpret the blueshift of this
\Pa\ emission as evidence for a net outflow along our sightline. The redshift
of the optical emission lines extracted from the nuclear aperture of the ESI
spectrum is consistent with our \Pa\ measurement.
}

{\new
We fit a Gaussian line profile to the \Pa\ emission in each spaxel of the data cube.
We then constructed  maps of the Doppler shift and velocity width of the line emission.
}
Fig.~\ref{fig:fullmaps} 
shows the results.
{\clm
Note that
the average velocity dispersion of the \Pa\ emission is
roughly 90\kms\ across the map in  Panel~{\it e} of  Figure~\ref{fig:fullmaps}. 
The typical linewidth is therefore roughly 200\kms\ FWHM and significantly
broader than the line spread function.
The  macroscopic velocity gradient is insufficient to broaden the lines this much, so the
ionized gas must have a significant turbulent component. 
}


\subsubsection{Circumnuclear Gas Kinematics}

{\clm 
In the circumnuclear region of Fig.~\ref{fig:model_bar}, the zero velocity line (ZVL) of the \Pa\ 
emission  runs northwest -- southeast at $PA = -46\deg$, but comparison to the continuum and \Pa\ 
contours in Fig.~\ref{fig:fullmaps}d  indicates the ZVL is offset southwest of the maximum 
infrared surface brightness. This offset persists if we 
adopt the \cite{Cicone2014} redshift; however, the ZVL moves towards the nucleus, 
closing about half the gap in Fig.~\ref{fig:fullmaps}d. 
A blueshift observed directly towards the nucleus, regardless of its exact size,  
is consistent with gas outflow on the near side of the circumnuclear disk.
It is also possible that the circumuclear disk is simply offset relative to 
the barycenter of the merger.  
}


{\clm
Another surprising feature of the ionized gas kinematics in the circumnuclear region
is the alignment between the velocity gradient  and the
{\it minor} axis of the infrared emission, which have the same position angle to 
within $10\deg$. Nuclear disks, in contrast, typically show a velocity gradient 
along their major axis.  The \Pa\ velocity dispersion in the circumnuclear region of IRAS~23365+3604
is approximately 110\kms, but the velocity along the major axis of the isophotes varies by less than
20\kms.  For comparison, \cite{Medling2014}  measured significantly higher ratios of circular to random
motion, $v/\sigma =  1 - 5$, in circumnuclear disks. If rotation dominates the circumnuclear velocity field 
in IRAS~23365+3604, then the underlying spheroid of ionized gas must be prolate rather than oblate like a disk.
}

{\clm
In Fig.~\ref{fig:model_bar}, we draw attention to two pairs of bends in the ZVL; these features
are robust to the exact choice for the systemic velocity. We interpret the coincidence of 
the inner bends and  the structure identified morphologically in Sec.~\ref{sec:data}
as evidence for gas inflow onto a circumnuclear disk or bar.
}
The ZVL  bends at the edge of a bar due to the distinctly non-circular motion of gas entering the ring 
\citep{Athanassoula1992a,Maciejewski2004a}. 
A comparison of the velocity field in the circumnuclear gas to Figures~6 and 9 of \cite{Davies2014} is instructive.
Those authors qualitatively fit the S20 bar model of \cite{Maciejewski2004b} to the velocity field of warm molecular 
gas in two active galaxies.  They find the ZVL nearly aligned with the minor axis of the isophotes.
In contrast, in the circumnuclear region of  IRAS 23365+3604, the ZVL is aligned with the major axis of the isophotes.
The bar interpretation therefore requires that the long axis of the bar is considerably foreshortened by 
projection on the sky; we estimate a bar diameter of  $ \approx\ 520 ~{\rm pc} / cos (36\deg)$, or 640~pc.  


\begin{figure}[h]
\centering
\includegraphics[width=0.85\linewidth, keepaspectratio=true, angle=0,clip=true, trim=200 0 0 0]{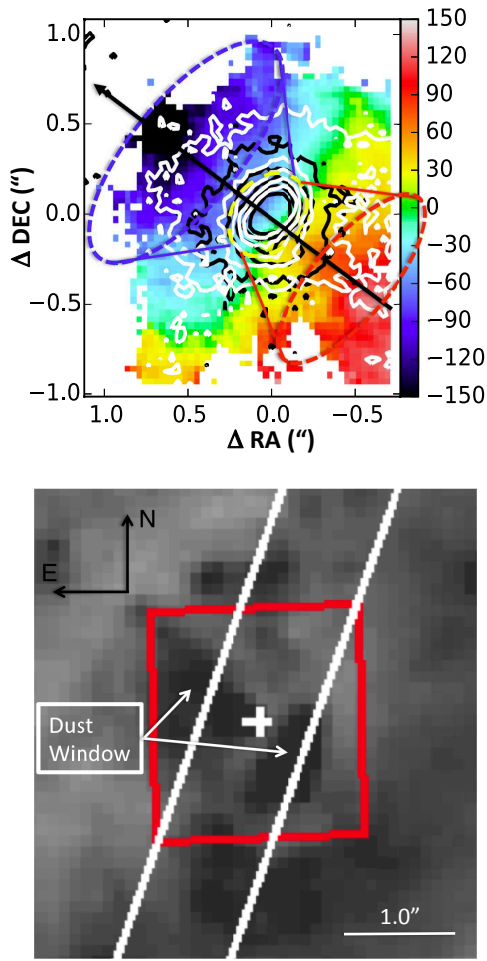}
\caption{
{\it Top:}
The \Pa\ velocity map with  a schematic representation of an outflow from the circumnuclear disk.  Contours are as defined
in Fig.~\ref{fig:fullmaps}, and the black arrow indicates the velocity gradient through the nucleus.
Feedback from young stars in the circumnuclear disk (yellow ellipse) drive the outflow into the 
approaching lobe (blue cone) and receding lobe (red cone).
The PA of the \Pa\ velocity gradient determines the axis of the bipolar flow.
The outflow cone is drawn with an opening angle $\theta = 45\deg$ and
include the ionized gas which blueshifted (redshifted)  more the 50\kms\ to the northeast (southwest).
The largest blueshift is measured  0\farcs79 (0.97 kpc) northeast of the nucleus. With the northeastern lobe of the bicone
tipped towards us, its  superposition on the circumnuclear disk produces a net blueshift of the ionized
gas towards the nucleus. 
{\it Bottom:}
Color map from Fig.~\ref{fig:zoom} zoomed for comparison to \Pa\ gas kinematics. We detect 
blue $B-I$ color in two regions of the OSIRIS map. These two {\it dust windows} lie along the \Pa\ 
velocity gradient. The outflow may have partially cleared out the dust here.
}
\label{fig:model_wind} \end{figure}


\subsubsection{Kiloparsec-Scale Gas Kinematics}

{\clm 
Fig.~\ref{fig:model_bar} shows how the ZVL bends again roughly 0\farcs8 from the nucleus. 
The symmetric shape with respect to the nucleus is again similar to the flow patterns induced
by bars. On this larger spatial scale, however, the infrared continuum is not deep enough
in the OSIRIS cube to identify a non-axisymmetric, stellar structure. Near the location of the 
outer ZVL bend, however, the $i$-band isophotes in Fig.~\ref{fig:fullmaps} twist from 
$PA \approx 50\deg$ (on smaller scales) towards the galactic major axis. The exact nature
of this non-axisymmetric stellar structure is not clear, but we suggest that it causes
the outer pair of bends in the ZVL. We believe these bends are the signature of gas
inflow from the larger scale molecular disk down to sub-kiloparsec scales.
}

Our OSIRIS map lies  within a single resolution element of \cite{Cicone2014}'s molecular gas map, 
so a direct comparison of the kinematics of the ionized and molecular gas is difficult. We simply
note that from the velocity channels in the radio data, they find a disk of molecular gas at 
$PA = -55\deg$ with the northwest side approaching (as indicated in Fig.~\ref{fig:fullmaps}a).
The line of nodes (LON), where the orbits intersect the plane of the sky,
for this disk of molecular gas is nearly perpendicular (to within $10\deg$) 
to the velocity gradient we see in the ionized gas at $PA = 44\deg$.  Hence the \Pa\ velocity
gradient is not in the direction expected from the rotation of the molecular gas disk.

%

\subsubsection{Interpretation: Superposition of Outflow and Inflow}

{\clm
The OSIRIS observation has resolved a kpc-scale structure of gas not previously
identified in IRAS~23365+3604. Considering the chaotic situation in the central kiloparsec of a merger,
the regular nature of this velocity gradient merits an explanation. Keeping in mind the considerably 
larger spatial scales of the molecular gas map, the kiloparsec-scale gas motion might reflect rotation
in a barred disk; the disk and bar scales being set by the CO (1-0) and F814W inner isophotes,
respectively. This interpretation would directly attribute the outer kinks in the ZVL to bar-induced
gas inflow, suggesting a doubly barred system \citep{Maciejewski2002,Maciejewski2004b}. 
We have several concerns, however, with this interpretation. First, the F814W isophotes do not have
the regularity of bars in normal spirals; and, second, this scenario places a circumnuclear disk/bar
within an outer bar that has an unusually small semimajor axis of just one kiloparsec.
In addition, the molecular gas is the dominant mass component in the central kiloparsec,
and it would have to  be highly concentrated to explain the velocity gradient with orbital
motion.}\footnote{
  \cite{Cicone2014} estimate a molecular gas   mass of $\log (M_{H_2}/\msun) = 9.93$. 
  Nearly all (90\%) of the CO(1-0) emission 
{\new
  comes 
}
from the central 5\farcs0 (6.20~kpc) diameter region. 
  Based on the nuclear K' magnitude from \cite{Surace2000}, the total stellar mass is of order  
  $9.8 \times 10^8$\msun\ in this region; an estimate which seems reasonable in comparison
  to mass measurements of other circumnuclear disks, i.e., $10^8$ to $10^{10}$\msun\ \citep{Medling2014}.  
  From the gravity of the molecular gas alone then, gas on
  circular orbits would reach a rotation speed of 110\kms\ at a radius of 3.1~kpc. 
  Our map of the ionized gas velocity field shows higher Doppler shifts, however,
  within a kiloparsec of the nucleus.}


{\clm
A galactic outflow offers an alternative explanation for the \Pa\ velocity gradient. 
The position angle of this velocity gradient suggests blowout perpendicular to 
the circumnuclear disk, i.e., along the path of least resistance. We sketch a schematic 
picture in Fig.~\ref{fig:model_wind}. Projection of the northeastern lobe of the outflow against
the circumnuclear disk also offers a plausible explanation for the blueshift of the ionized
gas measured towards the infrared nucleus and the increase in velocity dispersion northeast
of the nucleus.  Another advantage of the outflow interpretation is that it 
does not require the ionized gas to be in virial equilibrium. 
}

{\clm
We find additional evidence for a bipolar structure, in contrast to an axisymmetric disk, in the
dust distribution within the central kiloparsec of \i23365+3604. Two regions of reduced reddening 
are identified on opposite sides of the nucleus in the lower panel of Fig.~\ref{fig:model_wind}. 
These regions of bluer $B-I$ color are roughly a kiloparsec across. Their offsets  northeast and southwest 
of the nucleus are directly along the \Pa\ velocity gradient. A similar reduction in  reddening is seen 
in the circumnuclear regions of NGC 3227 and NGC~5643 where it has been attributed to an ionization cone 
produced by an outflow \citep{Davies2014}. The color map suggests the outflow cone has effectively 
displaced dust roughly one kiloparsec. 
}

The outflow interpretation does not explain all the features in the data cube.  It does not, for example,
explain the bends in the ZVL, which we still attribute to inflow on both the kiloparsec and circumnuclear
scales. 
{\new
Furthermore, while the wispy \Pa\ filaments, visible in Fig.~\ref{fig:fullmaps}b and discussed further in 
Sec.~\ref{sec:bar}, may turn out to be outflow features, we acknowledge that they do not line up closely with the cones sketched in Fig.~\ref{fig:model_wind}.
}
If the filaments are associated with an outflow, then the geometry is more complicated than the simple
cone depicted in Fig.~\ref{fig:model_wind}. 
{\clm 
We note some resemblances to Fig.~7 of \cite{Roskar2015} which shows a circumnuclear outflow in 
a simulation of a merger remnant; the gas flow is initially bipolar but takes on a more complicated
appearance before traveling even a full kiloparsec due to the chaotic nature of the surrounding 
interstellar medium. We suggest that the \Pa\ velocity field is the superposition of such an 
outflow and gas inflow towards the circumnuclear disk.}
Observations of AGN on similar spatial scales to our observation have shown combinations of molecular inflow and 
outflow superimposed on a rotating disk (\cite{Muller2011}, \cite{Hicks2013}, \cite{Rupke2013b},
\cite{Davies2014}, \cite{Medling2015}).



\section{Discussion}  \label{sec:discussion}

{\clm
Previous observations of IRAS~23365+3604 have shown that the merger has induced gas inflow
from large radii towards the central kiloparsec \citep{Soto2010} 
and that the starburst drives a galactic outflow which covers the entire 
galaxy \citep{Martin2006,Soto2012a}.
Analysis of the OSIRIS data cube reveals a circumnuclear disk or bar, the
kinematic signature of gas inflow towards this structure, and evidence for
a bipolar outflow emanating from it.
}
We further discuss these resolved gas flows and their relationship to galactic gas kinematics
in this section.

\subsection{Relationship of Nuclear Outflow to Galactic Wind}

The solid angle of the entire OSIRIS field of view is not much larger than a typical, seeing-limited spectroscopic aperture.
The echellete spectrum (obtained along the slit position shown in Fig.~\ref{fig:zoom}) shows two signatures of galactic outflow:
(1) a broad, blueshifted wing on the emission-line profiles and (2)  blueshifted resonance absorption. We plot these line
shapes relative to the systemic velocity in Fig.~\ref{fig:profile}. The spectra shown were extracted from a seeing-limited
observation of the nucleus; these outflow signatures have been mapped over several spatial resolution elements.

\subsubsection{Absence of Fast Outflow Detection in the Nucleus}

\begin{figure}[h] 

\centering
\includegraphics[width=0.8\linewidth, keepaspectratio=true, angle=-90, clip=true, trim = 0 0 0 0]{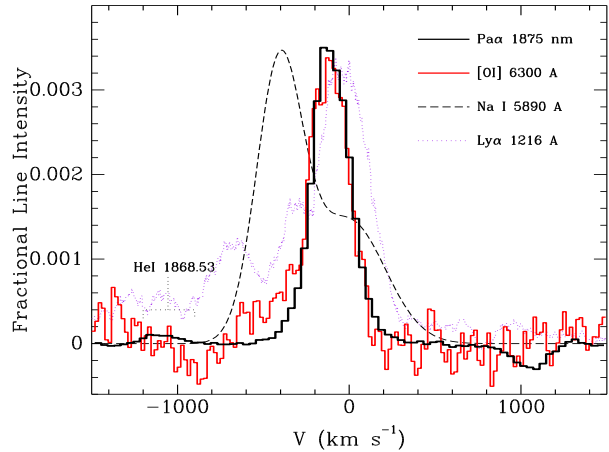}
\caption{
The integrated \Pa\ line profile from the entire OSIRIS map compared to the optical [OI] emission line profile.
The HeI1868.53~nm emission line is also detected in the line wing.
{\clm
The dashed line shows one component of 
the fit to the \naI\ doublet in a spectrum extracted from the nucleus; this is Aperture 3 
as described in \cite{Martin2006}.
The continuum level has been fit and subtracted in each case, and the resulting line profile normalized by the
central intensity as measured by a fitted Gaussian profile.
}
The [OI] profile is well fitted by two Gaussian components: (1) a narrow Gaussian ($\sigma = 81$\kms) with 
a small Doppler velocity of $V = -39$\kms, and (2) a blueshifted wing ($V = -290$\kms) which is much 
broader, $\sigma = 290$\kms \citep{Soto2012b}. The broad, blue wing on the [OI] emission profile 
is not detected in the OSIRIS data cube. The broad wing is most prominent in the \lya\ spectrum 
obtained through a 2\farcs5 diameter aperture  \citep{Martin2015}. 
The fitted \naI\ $\lambda 5890$ profile is shown inverted to facilitate comparison of the 
absorption to  the emission lines kinematics. The \naI\ shows a strong, blueshifted component 
at velocities similar to the blue wing on the optical 
emission-line profiles in addition to a component at the systemic velocity.
}
\label{fig:profile}  \end{figure}

To match apertures, we overplot the integrated \Pa\ line profile in Fig.~\ref{fig:profile}. 
The core of the \Pa\ and optical emission lines have 
the same Doppler shift and profile shape. The blueshifted wing on the optical lines extends to much larger Doppler shifts,
reaching $\approx -700$\kms, than does the \Pa\ line profile. The bulk Doppler shifts ($\pm 150\kms$) detected in the 
OSIRIS map are not nearly as  large as those of the blue wing on the optical line profiles. We do not detect the  
fast outflow marked by broad, blueshifted component in optical spectra in the OSIRIS data cube. 
The broad line components in the optical spectrum are 
almost exclusively shock excited \citep{Soto2012b}. 
In contrast, the narrow component, which appears to be directly 
associated with the \Pa\ profile, has line ratios indicative of photoionization by massive stars.

When we see deeper into the ULIRG in the near-infrared, this line wing turns out to be 
much less prominent than in the optical.  This result is difficult to understand unless the narrow
component is indeed more reddened than the highly blueshifted gas; in other words, the line wing
becomes weaker relative to the core when we can probe deeper into the ULIRG. Based on its absence in
the \Pa\ map, the highly blueshifted gas is not coming from the unresolved nucleus.

\cite{Martin2015} proposed that the fast outflow was recently part of the hot wind fluid. 
The higher density in ULIRG winds, or any starburst with a high SFR surface density, allows
the wind to cool down to the inflection in the cooling function around $10^{7.2}$~K where
thermal instability leads to condensations within the hot wind. In this physical picture,
the molecular component of the outflow  \citep{Veilleux2013,Cicone2014}
may also form in situ in the fast outflows.

The Herschel-PACS observation of IRAS~23365+3604 shows a strong P-Cygni profile in the  OH~119\um\ $+ ^{18}$OH~120\um\  
complex. The blueshifted absorption troughs reach Doppler shifts up to -1300\kms \citep{Veilleux2013}.  
The OH line profile integrates the signal from a $9\arcsec \times 9\arcsec$ spatial region, however, so we re-examine
the CO (1-0) map of \cite{Cicone2014} for clues about the physical location of the high velocity outflow. The
10\% of the line flux not modeled by the Gaussian fit lies in low intensity line wings; detections
in the channel maps over $-600 < V_{CO}(\kms)  < -300$ show the highest velocity CO emission comes from a region 
east of the nucleus just beyond the region mapped with OSIRIS. 

Resonance absorption from \naI\ must trace gas sheltered from the ionizing continuum, so we might expect
some rough spatial association with the molecular outflow. Since the optical transition is a doublet,
we show the  fitted profile of only the $\lambda 5896$ transition in Fig.~\ref{fig:profile} to avoid confusion from the blend.
The systemic absorption covers a velocity range slightly broader than the narrow emission component. The
velocity range of the outflow component matches that of the blue wing seen on the optical emission lines.

\cite{Martin2006} mapped the \naI\ absorption across the ULIRG 
demonstrating that the outflow covers the galaxy and is not confined to the nuclear region. In their Fig.~1p,
a spectrum of the nucleus (Aperture 3) shows a blueshifted component substantially  stronger than the absorption
at the redshift of the galaxy. Moving along the slit to the north, the systemic component grows in strength
relative to the outflow component, whereas the outflow component continues to dominate the absorption trough
towards the south (e.g., aperture 2). This velocity shift of the trough along the slit is smaller, and in the
opposite direction to, the gradient in the \Ha\ Doppler shift as can be seen Fig.~2o of \cite{Martin2006}.
{\clm
The prominence of the \naI\ outflow to the south-southeast (along the ESI slit) is consistent with the low-ionization-state
 outflow being coincident with the CO outflow, which is mapped in Fig.~3c of \cite{Cicone2014}.
}

\subsubsection{Properties of The Nuclear Outflow}

We can relate the gas kinematics in the \Pa\ map to the structure seen in an \Ha\ spectrogram, which is shown
in Figure~1p of \cite{Martin2006}. Within $\pm 3\arcsec\ $ of the nucleus along the slit, the \Ha\ is
blueshifted (redshifted) to the northwest (southeast).  Although \cite{Martin2006} attributed this velocity 
shear as a combination of orbital and rotational motion in merging gas disks, the higher resolution view 
afforded by OSIRIS does not resolve two nuclei thereby ruling out this interpretation.  The \Pa\  map reveals 
the position angle of the velocity gradient is nearly perpendicular to that of the molecular disk. The shear in the
\Ha\ spectrum  is consistent with the \Pa\ map at the same $PA$.

The velocity gradient seen in our \Pa\ map contributes to the width of the \Pa\
line in the circumnuclear spectrum. The blueshifts (and redshifts) detected in the northeast 
(southwest) corner of the OSIRIS data cube are slightly larger than the velocity dispersion 
towards the brighter circumnuclear disk/bar. The effect is not large because the most Doppler
shifted emission regions have relativley low surface brightness. The bulk motion, however,
would clearly contribute to the linewidth in seeing-limited, spectroscopy of the 
circumnuclear region.

{\clm
We expect to see gas outflow from the nucleus considering its well established presence on 
large spatial scales, so we briefly examine the consequences of such an interpretation.
To the northeast, the \Pa\ is blueshifted up to -150\kms, and the Doppler shift reaches $+150$\kms\ 
towards the southwest corner of the OSIRIS map. A model with a constant velocity outflow, in contrast, produces radial
isovelocity contours. A successful outflow model must account for this velocity gradient along 
the axis of the outflow cone. Acceleration over spatial scales from about 100 pc up to a kiloparsec
could be produced by the radiation pressure from a central concentration of stars as described by
\cite{Murray2011}; alternatively, velocity segregation in the flow would lead to higher velocity gas
at larger radii (because it moves more quickly than low velocity gas).   
}

In principle, for a particular parameterization of the radial velocity, one could fit the
inclination and opening angle of the cone. Here, we   estimate the outflow cone is tipped roughly 47\deg\ 
relative to our sightline based simply on the axis ratio of the stellar nuclear disk. If we assume an intrinsically
round  disk for illustration, then the  inclination $i = 43\deg$, and we infer a radial outflow speed of 200\kms\ for the ionized gas.

\subsection{The Formation of a Circumnuclear Disk: }  \label{sec:bar}

\begin{figure*}
\centering
\includegraphics[width=0.75\linewidth, keepaspectratio=true, angle=90, clip=true, trim = 100 0 0 30]{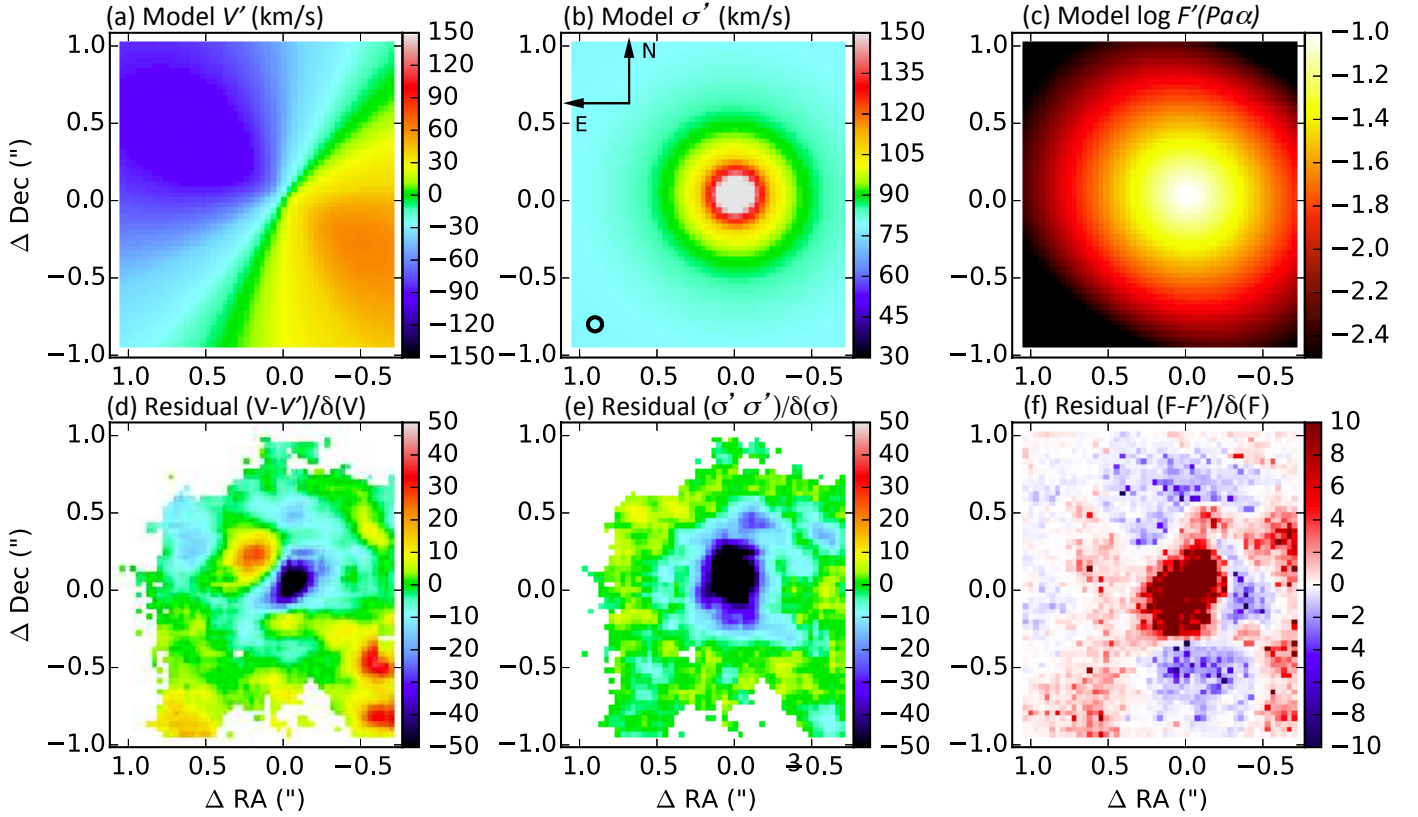}
\caption{
{\clm 
The velocity field, velocity dispersion, and surface brightness for a fitted disk 
(a-c) and the fit residuals as a fraction of the error (e-f). The spatial resolution of the data is indicated
by the small circle in the lower, left corner of (b).
The disk model adopts an exponential function to describe the radial decline in surface brightness.
The surface brightness profile perpendicular to the plane of the disk scales as ${\Sigma(z) \propto {\rm exp}(-0.5z^2/h_z^2)}$, 
where the vertical thickness of this disk was fixed at a constant fraction of the half-light radius such that
$h_z  = 0.2 R_{1/2}$.  The line-of-sight velocity dispersion includes several terms, described in Table~\ref{tab:disk}, to 
simulate the observed total velocity dispersion $\sigma_{\rm tot}$. Line profile comparisons are shown in Fig.~\ref{fig:fullmaps}.
}
%
%
%
}
\label{fig:residuals} \end{figure*}

{\clm 
Our infrared observations resolve a non-axisymmetric, circumnuclear structure.
}
Its size lies within the range of the nuclear disks identified recently by \cite{Medling2014} 
in LIRGs and ULIRGs. The structure of such disks is shaped largely by radiation
pressure \citep{Scoville2003},  and models including this process predict 
two classes of nuclear disks: (1) those with a starburst on large scales that consumes all of the gas with 
little or no fueling of a central AGN and (2) those with an outer large scale 
starburst accompanied by a more compact starburst on 1-10 pc scales and a bright 
central AGN \cite{Thompson2005}.  The circumnuclear disk in IRAS~23365+3604 appears to still be
in the former state, and the cool infrared color of this ULIRG is consistent with
the spatially extended starburst powering the majority of the bolometric luminosity.  

We find signs of gas inflow onto this structure. The bends in the ZVL near this structure
reveal deviations from circular motion, and similar features are expectated from
the radial infall produced by nuclear bars and spirals. However, whereas \cite{Medling2014} 
observe rotation about the minor axis of the nuclear disks, the ZVL of our \Pa\ map
follows the major axis of the circumnuclear structure. 
{\clm
In order to produce this
offset between the LON and the major axis, a model of the circumnuclear structure in 
\i23365+3604 would appear to require an intrinsically non-axisymmetric component. 
}

{\clm
To better understand the nature of this circumnuclear structure, we considered the
gas kinematical signatures of nuclear bars and  spirals. For example, as gas responds to a non-axisymmetric 
potential, a nuclear spiral forms that may either be damped leading to the formation of a nuclear 
ring or get strengthened and 
{\new
propagate
}
 towards the SMBH as a spiral shock \cite{Maciejewski2004b}.
} 
In the models shown in Figures 6 and 9 of \cite{Davies2014}, the LON within the ring is along
the major axis, and the ZVL is along the minor axis within the ring (opposite to what
we see in IRAS 23365+3604). The ZVL bends abruptly at the ring due to the mostly radial inflow along
the spiral arms at larger radii. To fit a similar model to our data cube, where the major axis
is not along the LON, projection on the sky must significantly foreshorten the major axis. Hence,
the circumnuclear structure would be intrinsically even less round than it appears in projection.



{\new
To look more carefully for signatures of spiral shocks feeding the circumnuclear disk, 
we fit a simple, disk model to  the data cube.
}
We used the GalPak~$^{\rm 3D}$ code recently introduced
by \cite{Bouche:2015} and  fit the intensity of the \Pa\ emission in each voxel. 
GalPak$^{\rm 3D}$ convolves an intrinsic model with a 3D kernel
and returns a model data cube. 
We adjusted this kernel to describe the line spread function (LSF) and PSF
of the OSIRIS observation. 
The LSF convolved with the combined spectra in the modeled 
disk produces emission lines at the instrumental spectral resolution of $\Delta V \approx 100$\kms.
The code uses a Markov Chain to efficiently explore parameter space. Uncertainties in the fitted 
parameters are estimated from the posterior probability distribution after 20,000 Monte Carlo iterations. 
We list the fitted parameters and their values in Table~\ref{tab:disk}.

%
%



We subtracted this model from the data to highlight the non-axisymmetric structures
as illustrated in Fig.~\ref{fig:residuals}. The circumnuclear disk dominates the
residuals. The flux residuals in Panel~{\it f} also reveal three filaments running 
approximately 1 kpc to the edge of the OSIRIS map and a fourth, shorter filament to 
the northeast.   These filaments do not resemble spiral arms. The southwest and southeast 
filaments, if they were arms, would have very different pitch angles than would the western filament. 
Neither do we recognize a nuclear spiral in the color map shown in Fig.~\ref{fig:model_wind},
whereas the nuclear spirals often found in the central kpc of active galaxies 
imprint a recognizable signature in color maps \citep{Martini1999,Davies2014}.  

Velocity residuals offer another method for identifying nuclear spirals through a pattern of 
radial inflow and outflow in the plane of the disk \citep{Davies2009}. The velocity residuals
in Panel~{\it d}, however, show no correlations with the locations of the \Pa\ filaments.
The largest residuals correspond to the bluest regions in the $B-I$ color map in 
Fig.~\ref{fig:model_wind}, and we attribute this correspondence to an outflow.

Finally, we would also expect local increases in velocity dispersion from the shocks 
associated with a bar or spiral shock. Because the infall maintains some angular momentum, 
it hits dense gas in the arm on the opposite side (Davies et al. 2009). The velocity dispersion
residuals in Panel~{\it e}, however, are largely axisymmetric. Figure~\ref{fig:fullmaps}f
shows that the line profiles from the disk model are too broad in the nucleus and too narrow at 
large radii.

{\clm
We conclude that the circumnuclear structure in \i23365+3604 although the size of a typical
disk shows unusual gas kinematics which suggests it is not yet an axisymmetric disk. This
ULIRG may be observed soon enough after the coalescence of the two cores that the 
circumnuclear disk has not completely reformed. Figure~9 of \cite{Roskar2015}, for example,
shows that a merger remnant can require up to 10~Myr to rebuild a regular disk. Those
simulations predict that such structures contain two SMBHs separated by several tens of
parsecs.  Considering the inflow timescales measured for this galaxy on scales of several
kiloparsecs \citep{Soto2010}, we are observing \i23365+3604 roughly 100~Myr after the
final passage.
}


\begin{deluxetable*}{l c c c c c c }
\tablewidth{0pt}
\tablecaption{ \label{tab:disk} Fitted Parameters of Disk Models}
\tabletypesize{\small}
\tablehead{
(1)	& (2)	& (3)	& (4)	& (5)	& (6) & (7) \\
Disk 		&${\rm R_d}$&	$i$	&	PA	&		${\rm V_{max}}$	&	$\sigma_i$	&	$\chi_{\nu}$ \\
Model       & (\arcsec)	& (\deg) & (\deg)		& (\kms) & (\kms) &  
}
\startdata
Unmasked Mass 		& 0.217  $\pm$  0.001 & 22.1  $\pm$  0.1 & 55.5  $\pm$  0.1   			     & 197.8  $\pm$  0.5    & 73.1  $\pm$  0.1 &  4.64 \\
Masked Mass 		& 0.270  $\pm$  0.001 & 20.0  $\pm$  0.1 	& 55.4  $\pm$  0.1  			      		& 243.1  $\pm$  0.8   & 71.5  $\pm$  0.2    & 3.88 
\enddata
\tablecomments{Col.(1): We experimented with masking out the central region of the data cube and fitting only the large-scale 
velocity gradient.  The   masking improves the fit statistic  but produces little change in best-fit disk parameters. 
Col.(2): Exponential radial scalelength. The radial scalelengths become systematically larger when the central \Pa\ emission 
is masked, entirely consistent with the flattening of the \Pa\ surface brightness profile described in Sec.~\ref{sec:morph_gas}.
Col.(3): Disk inclination varies only slightly among models.
Col.(4): Disk position angle (measured going east from north) varies only slightly among models. 
Col.(5):  Asymptotic rotation speed computed using    
the {\it mass model} defined by $V(R) = \sqrt{{\rm G~M}_{enc}(R) / R} $, where $\rm M_{enc}(R)$ is the mass 
within radius, $R$. This definition ties the velocity profile to the chosen surface brightness profile. In the model, 
$\rm M_{enc}(R)$ is determined numerically by summing the flux as a function of radius and assuming 
that the ratio of enclosed mass to \Pa\ flux is constant with radius. This profile is then normalized to its 
maximum within the region modeled.  
Col.(6):
The intrinsic velocity  dispersion, $\sigma_i$, of the gas which accounts for the internal gas dynamics of the disk.
The total velocity dispersion is defined as $\sigma_{\rm tot} = \sqrt{\sigma_d^2 + \sigma_i^2 + \sigma_m^2}$,
which includes terms for mixing, $\sigma_m$, along the line of sight (appropriate to a geometrically 
thick disk) and the velocity dispersion produced by the disk self-gravity, $\sigma_{d}/h_z \equiv V(r) / r$.
Col.(7): Reduced chi-squared fit statistic.}
%
\end{deluxetable*}

\section{Summary}  \label{sec:summary}

{\clm
The spatial resolution delivered by the Keck AO system resolves a young, circumnuclear disk in IRAS~23365+3604.
Its radial scalelength is  $R_e = 0\farcs130$ (161~pc) in the starlight and ionized gas, and its position angle on the sky 
happens to}
nearly coincide with that of a molecular disk identified previously on much larger spatial scales (and at much lower 
spatial resolution) \citep{Cicone2014}.
The isophotes of the ionized gas are rounder than the starlight and extend into filaments traceable over a 
kiloparsec to the edges of the \Pa\ map .

The \Pa\ line is blueshifted northeast of this circumnuclear disk and redshifted to the southwest.
The large-scale velocity gradient of roughly 150\kms\ per kiloparsec establishes
the source of the shear previously observed in the \Ha\ spectrum
near the nucleus \cite{Martin2006} and broadens the core of the photoionized line component in the circumnuclear spectrum.
Remarkably, however, this gradient is nearly perpendicular to the major axis of both the infrared continuum 
and CO (1-0) isophotes, so it cannot mark the rotation of a oblate disk of ionized gas.

The isovelocity contours in the OSIRIS \Pa\ map show significant deviations from regular motion. In
particular, two sets of symmetric bends in the ZVL provide  evidence for gas inflow into the circumnuclear 
region. The outer bends plausibly mark
inflow into a gas bar previously unresolved in the CO (1-0) observation. 
Within this kiloparsec scale disk/bar, we found both kinematic and morphological evidence 
for a nuclear bar or irrgular disk.

The absence of a highly blueshifted wing on the integrated \Pa\ line profile,
or anywhere within the field mapped with OSIRIS, poses an interesting paradox.
A prominent wing is visible on both the H Balmer lines and the optical forbidden
lines in an ESI spectrum of the nucleus; and  this high velocity outflow is shock
excited \citep{Soto2012a}. Qualitatively, the high velocity outflow becomes
more prominent relative to the line core at shorter wavelengths. It is most
prominent in \lya\ emission for example, where the core is likely depressed
due to resonance scattering but the line wings appear to be direct emission
\citep{Martin2015}. One possible explanation is that because we see a much
larger volume of the nucleus in the near-infrared (due to the reduced extinction),
the increased strength of the  narrow line core leaves the emission in the line
wing (potentially emitted by gas at much larger radii) much less prominent. 

{\clm
The starburst fueled by the galaxy merger over the past 100~Myr drives a previously
studied global outflow. The \Pa\ velocity gradient in the OSIRIS map provides
evidence for a circumnuclear outflow emanating from a region where the circumnuclear
disk is forming following the final coalescence of the galactic cores.
Factors favoring this interpretation of the \Pa\ emission include (1) the orientation of the \Pa\ velocity gradient, which 
is consistent with a  circumnuclear outflow collimated by the disk of molecular gas, (2) the net blueshift 
of the ionized gas emission towards the nucleus, and (3) the spatial correlation between the
the steepest velocity gradient and the minimum reddening.  For an  outflow that dominates the \Pa\ emission, 
we infer a launch region the size of the circumnuclear disk (radius 520~pc) and estimate an
acceleration of roughly 150\kms\ over one kiloparsec. Calculations for outflows
from massive star clusters indicate the radiation pressure from just a few million
\msun\ cluster could easily produce this acceleration, see Figure 4 of \cite{Murray2011}.
}

{\clm
Alternatively, the \Pa\ velocity gradient is due to gas orbiting in a bar-like structure
roughly two kiloparsecs across. The position angle of the  inner F814W isophotes
offer a plausible, but not compelling, stellar component.  The existing CO (1-0) observations lack
the resolution to resolve even this kiloparsec-scale, non-axisymmetric structure.
In the future, we hope to resolve the kinematics of the warm, molecular gas
in $H_2 (1-0)$ emission and  elucidate the nature of the overall velocity gradient by 
resolving the morphology of the shocked gas in this dynamic environment.
}


%

Our results demonstrate that gas dissipation is a very important process for forming nuclear structures.
We show that the patches of lower reddening in the circumnuclear region are spatially coincident with the 
circumnuclear outflow. We have caught the outflow sweeping dust away from the nuclear region where the
nascent AGN remains highly obscured in this system. 
{\clm
The irregular structure of the circumnuclear disk may reflect its youth,  and we suggest
\i23365+3604 would be an excellent target in which to look for a pair of SMBHs separated by a few tens of parsecs.
}

\acknowledgments{
This research was partially supported by the National Science Foundation under AST-1109288.
We thank Nicolas Bouch\'{e}, Claudia Cicone, Eric Emsellem, and Roberto Maiolino for stimulating discussions that improved this manuscript.
Alexander Adams  contributed preliminary modeling which appeared in Soto (2012) and influenced the direction
of our analysis. A portion of this manuscript was written while CLM visited
the Aspen Center for Physics which is supported by the National Science Foundation under Grant No. NSF PHYS-1066293.
We also acknowledge the very significant cultural role and reverence that the summit of Mauna Kea has always had within the 
the indigenous Hawaiian community. We are most fortunate to have the opportunity to conduct 
observations from this mountain.}

\noindent
{\it Facilities:}  \facility{Keck}

\clearpage
\bibliographystyle{myapj}
\bibliography{library}

\end{document}